\documentclass[
seceq,
preprint]{ptptex}

\usepackage{amssymb}

\newcommand{\Dslash}{{\not}\kern-0.05em D}
\newcommand{\dslash}{{\not}\kern+0.1em\partial}
\newcommand{\nablaslash}{{\not}\kern+0.05em\nabla}

\newcommand{\tr}{\mathop{\rm tr}\nolimits}

\newcommand{\Pf}{\mathop{\rm Pf}\nolimits}

\newcommand{\SO}{\mathop{\rm SO}}
\newcommand{\U}{\mathop{\rm {}U}}
\newcommand{\rmd}{{\rm d}}
\newcommand{\rmD}{{\rm D}}
\newcommand{\ring}{\mathaccent"7017 }

\newcommand\fverb{\setbox\pippobox=\hbox\bgroup\verb}
\newcommand\fverbdo{\egroup\medskip\noindent%
                        \fbox{\unhbox\pippobox}\ }
\newcommand\fverbit{\egroup\item[\fbox{\unhbox\pippobox}]}
\newbox\pippobox

\preprintnumber[3cm]{
RIKEN-TH-68\\NIIG-DP-06-1
\\{\tt hep-lat/0604003}
}

\title{
Overlap lattice fermion in a gravitational field%
}

\author{
Masashi \textsc{Hayakawa},$^{1,}$\footnote{E-mail: haya@riken.jp}
Hiroto \textsc{So}$^{2,}$\footnote{E-mail: so@muse.sc.niigata-u.ac.jp}
and Hiroshi \textsc{Suzuki}$^{3,}$\footnote{E-mail: hsuzuki@riken.jp}
}

\inst{
$^{1,3}$Theoretical Physics Laboratory, RIKEN, Wako 2-1, Saitama 351-0198,
Japan\\
$^2$Department of Physics, Niigata University, Ikarashi 2-8050, Niigata,
950-2181, Japan
}

\abst{
We construct a lattice Dirac operator of overlap type that
describes the propagation of a Dirac fermion in an external gravitational
field. The local Lorentz symmetry is manifestly realized as a lattice gauge
symmetry, while it is believed that the general coordinate invariance is
restored only in the continuum limit. Our doubler-free Dirac operator satisfies
the conventional Ginsparg-Wilson relation and possesses $\gamma_5$ hermiticity
with respect to the inner product, which is suggested by the general
coordinate invariance. The lattice index theorem in the presence of a
gravitational field holds, and the classical continuum limit of the index
density reproduces the Dirac genus. Reduction to a single Majorana fermion
is possible for $8k+2$ and $8k+4$ dimensions, but not for $8k$ dimensions,
which is consistent with the existence of the global gravitational/gauge
anomalies in $8k$~dimensions. Other Lorentz representations, such as
the spinor-vector and the bi-spinor representations, can also be treated.
Matter fields with a definite chirality (with respect to the
lattice-modified chiral matrix) are briefly considered.
}

\begin{document}

\maketitle

\section{Introduction}
It is now well-known that a Dirac operator defined on a Euclidean lattice
allows an exact chiral symmetry of a lattice-modified
form~\cite{Luscher:1998pq} if the lattice Dirac operator satisfies the
Ginsparg-Wilson relation~\cite{Ginsparg:1982bj,Hasenfratz:1998ft}. Such
Dirac operators in lattice gauge theory were discovered in the context of
the perfect action approach~\cite{Hasenfratz:1998ft,Hasenfratz:1998jp} and
in the context of the overlap formulation~\cite{Neuberger:1998fp}. The
lattice action of a massless Dirac fermion is exactly invariant under the
modified lattice chiral transformation, and the functional integration
measure is invariant under flavor non-singlet chiral transformations. Under
a flavor-singlet $\U(1)$ chiral rotation, the integration measure is not
invariant and gives rise to a non-trivial Jacobian, just as in the Fujikawa
method~\cite{Fujikawa:1979ay} in the continuum theory; the continuum limit
of the Jacobian reproduces the axial $\U(1)$
anomaly~\cite{Kikukawa:1998pd,Fujikawa:1998if,Adams:1998eg,Suzuki:1998yz,%
Fujiwara:2002xh} (see also Ref.~\citen{Chiu:1998qv}). Moreover,
the axial anomaly obtained from the Jacobian possesses a
topological property and implies the lattice index theorem. (We reproduce the
argument demonstrating this point by using our lattice Dirac operator for the
gravitational interaction.) This topological property of the axial anomaly
allows a cohomological analysis of the anomaly~\cite{Luscher:1999kn} with
finite lattice spacings, and this analysis is crucial in the formulation of
lattice chiral gauge theories~\cite{Luscher:1999du}. See
Refs.~\citen{Kaplan:1992bt} and~\citen{Niedermayer:1998bi} for related works
and a review of these developments.

The above describes the situation in lattice {\it gauge\/} theory. Now, it is
natural to ask to what extent the above scenario holds in the presence of
a {\it gravitational\/} field. In the continuum theory, it is well-known
that the chiral symmetry suffers from the quantum anomaly also in the
presence of a gravitational field~\cite{Kimura:1969}, and this anomaly has a
topological meaning expressed by the index
theorem~\cite{Atiyah:1967,Hawking:1979zs,Alvarez-Gaume:1983at}. The purpose of
this paper is to determine how these points can be naturally understood in a
framework of lattice field theory.

In this paper, we construct a lattice Dirac operator of overlap
type~\cite{Neuberger:1998fp} that describes the propagation of a Dirac fermion
in an external gravitational field. With suitable modifications for curved
space, our lattice Dirac operator satisfies the Ginsparg-Wilson relation in its
conventional form and possesses $\gamma_5$ hermiticity. These properties imply
the validity of the lattice index theorem and a topological
property of the axial $\U(1)$ anomaly. Our formulation offers, in a
well-defined lattice framework, a conceptually simple way to understand the
axial anomaly in the presence of a gravitational field.

Here we do not intend to study the non-perturbative dynamics of gravity on
a spacetime lattice. We also do not consider a possible non-trivial global
topology of spacetime. These problems are beyond the scope of this paper.
We believe, however, that studying the interaction of a lattice Dirac field
to an external gravitational field will be useful in the investigation of the
coupling of matter fields to dynamically discretized gravity. For previous
studies of the gravitational interaction of matter fields defined on a
lattice, see Ref.~\citen{Sorkin:1975ah} and references therein.

Our formulation is a natural generalization of the overlap-Dirac operator in
lattice gauge theory to a system coupled to external gravitational
fields. As easily imagined, the local Lorentz symmetry in curved space can
be manifestly realized as an internal gauge symmetry on the lattice, while
the general coordinate invariance is not manifest, because our base space
takes the form of a conventional hypercubic lattice. With regard to the latter
point, our goal is modest, seeking only the situation in which the general
coordinate invariance is restored in the limit $a\to0$, where $a$ denotes the
lattice spacing.

\section{Preliminaries}
We consider a $2n$-dimensional hypercubic lattice. We regard it as a
discretized approximation of a family of coordinate curves on a Riemannian
manifold with Euclidean signature.
(We consider only torsion-free cases in this paper.) In particular, $4n$
links emanate from each of the lattice sites.\footnote{In our present
formulation, in which the vielbein~$e_\mu^a(x)$ exists on each lattice site~$x$,
it is natural to regard the index~$\mu$ of $e_\mu^a(x)$ as labeling the
directions of the links starting from the site~$x$. Then, if the Lorentz
index $a$ runs from $0$ to $2n-1$, the index~$\mu$ should also run from $0$ to
$2n-1$ for the matrix~$e_\mu^a(x)$ to be invertible. In this sense, the
restriction to a hypercubic lattice is natural.} In this
sense, our lattice is so primitive that it is not directly related to
the Regge calculus or to dynamical triangulation.

The spacetime indices are denoted by Greek letters, $\mu$, $\nu$, \dots, and
run from $0$ to $2n-1$, and the local Lorentz indices are denoted by Roman
letters, $a$, $b$, \dots, and run also from $0$ to $2n-1$. Summation over
repeated indices is implied for expressions in the continuum theory, while for
expressions in the lattice theory, we always explicitly indicate a summation
over indices. The unit vector along the $\mu$-direction, for example, is
denoted by~$\hat\mu$. In most of expressions below, we set the lattice spacing
to unity (i.e., $a=1$) for simplicity. The Euclidean gamma matrices
satisfy the relations\footnote{Here we intentionally use a combination of the
complex conjugation symbol~$*$ and the transpose operation symbol~$T$ with
respect to the spinor indices instead of the dagger~$\dagger$, because we want
to use the symbol~$\dagger$ to represent hermitian conjugation with respect to
the inner product~(\ref{threexthree}). This avoids possible confusion concerning
the chirality constraint in curved space (see \S5).}
\begin{equation}
   \{\gamma^a,\gamma^b\}=2\delta^{ab},\qquad
   (\gamma^a)^{T*}=\gamma^a.
\label{twoxone}
\end{equation}
The chiral matrix and the generator of $\SO(2n)$ in the spinor
representation are defined by
\begin{eqnarray}
   &&\gamma_5\equiv i^n\gamma^0\gamma^1\cdots\gamma^{2n-1},\qquad
   (\gamma_5)^{T*}=\gamma_5,
\nonumber\\
   &&\sigma^{ab}\equiv{1\over4}[\gamma^a,\gamma^b].
\label{twoxtwo}
\end{eqnarray}

In addition to the lattice Dirac fermion field $\psi(x)$, the vielbein
$e_\mu^a(x)$ is also defined on the lattice sites.
The determinant of the vielbein is denoted by
\begin{equation}
   e(x)\equiv\det_{\mu,a}\{e_\mu^a(x)\},
\label{twoxthree}
\end{equation}
and it is assumed to satisfy
\begin{equation}
   e(x)>0,\qquad\hbox{for all $x$}
\label{twoxfour}
\end{equation}
as in the continuum theory. The inverse matrix of the vielbein is denoted by
$e_a^\mu(x)$.

As a lattice counterpart of the spin connection, we introduce gauge
variables on the lattice links:
\begin{equation}
   U(x,\mu)\in\mathop{\rm spin}(2n).
\label{twoxfive}
\end{equation}
In this paper, the vielbein $e_\mu^a(x)$ and the link variables $U(x,\mu)$
are treated as external non-dynamical fields. They are, however, not
independent of each other and are subject to a certain constraint, which is
analogous to the metric condition. This constraint is explored in the
next subsection. We often use the abbreviated notation
\begin{equation}
   \gamma^\mu(x)\equiv\sum_a e^\mu_a(x)\gamma^a.
\label{twoxsix}
\end{equation}
Note that $\gamma^a$ and $\gamma^\mu(x)$ do not commute with the link
variables~$U(x,\mu)$, i.e., $[\gamma^\mu(x),U(y,\nu)]\neq0$.

The local Lorentz transformation has a natural realization on the lattice
as follows:
\begin{eqnarray}
   &&\psi(x)\to g(x)\psi(x),\qquad
   \overline\psi(x)\to\overline\psi(x)g(x)^{-1},
\nonumber\\
   &&U(x,\mu)\to g(x)U(x,\mu)g(x+\hat\mu)^{-1},
\nonumber\\
   &&\gamma^\mu(x)\to g(x)\gamma^\mu(x)g(x)^{-1},\qquad
   e_a^\mu(x)\to{1\over2^n}\sum_b
   \tr\{\gamma_a g(x)\gamma^bg(x)^{-1}\}e_b^\mu(x),
\label{twoxseven}
\end{eqnarray}
where the gauge transformation $g(x)\in\mathop{\rm spin}(2n)$ is defined
on each site~$x$. Note that $e(x)$ is invariant under local Lorentz
transformations. Covariance under this transformation is manifest in our
construction, exactly as in the case of an internal gauge symmetry in
lattice gauge theory. Contrastingly, invariance or covariance under general
coordinate transformations is very difficult to implement in a lattice
spacetime. We conjecture that this is restored only in the continuum limit.
Below, we examine whether the correct index in the continuum limit is
reproduced as the classical continuum limit, $a\to0$,
of the lattice chiral anomaly. Our result indicates that the violation of
general coordinate invariance is $O(a)$ in our quantum theory of fermion
fields coupled to non-dynamical gravitational fields.

\section{Overlap-Dirac operator for the Lorentz spinor field}
\subsection{Inner product, hermiticity and the metric condition}
We start our analysis with a lattice Dirac operator which is defined in terms
of the nearest-neighbor forward covariant difference:
\begin{equation}
   \nablaslash\psi(x)\equiv\sum_\mu
   \gamma^\mu(x)\{U(x,\mu)\psi(x+\hat\mu)-\psi(x)\}.
\label{threexone}
\end{equation}
This clearly behaves covariantly, i.e.,
$\nablaslash\psi(x)\to g(x)\nablaslash\psi(x)$, under the local Lorentz
transformation~(\ref{twoxseven}). Then, setting
\begin{equation}
   U(x,\mu)={\mathcal P}\exp\left\{\int_0^1\rmd t\,
   {1\over2}\sum_{ab}\omega_{\mu ab}(x+(1-t)\hat\mu)\sigma^{ab}\right\},
\label{threextwo}
\end{equation}
with a smooth spin connection field $\omega_{\mu ab}(x)$, the {\it naive\/}
continuum limit of $\nablaslash$ coincides with the Dirac operator in the
continuum, denoted by $\Dslash$. More precisely, we have
$\lim_{a\to0}\nablaslash\psi(x)=\Dslash\psi(x)$ for any fields~$\psi(x)$,
$e_\mu^a(x)$ and $\omega_{\mu ab}(x)$ which vary slowly over the
lengthscale~$a$.

Now to ensure that the lattice index theorem holds in lattice gauge theory,
in addition to the Ginsparg-Wilson relation, the $\gamma_5$ hermiticity of a
lattice Dirac operator is very important~\cite{Niedermayer:1998bi}. We are
therefore naturally led to attempt to clarify the meaning of hermiticity in the
presence of a gravitational field, because the concept of hermiticity involves
the definition of the inner product. In the continuum theory, the inner product
is modified in curved space for general coordinate invariance.

We introduce the inner product of two functions on the lattice with spinor
indices as
\begin{equation}
   (f,g)\equiv\sum_xe(x)f(x)^{T*}g(x),
\label{threexthree}
\end{equation}
by using the determinant of the vielbein, $e(x)$. The inner
product~(\ref{threexthree}) is a natural lattice counterpart of the general
coordinate invariant inner product in curved space, and the norm of any
non-zero function with respect to this inner product is positive definite. We
believe that inclusion of the vielbein~$e(x)$ is important for the restoration
of general coordinate invariance in the continuum limit.

We next examine
\begin{eqnarray}
   &&(f,\nablaslash g)
\nonumber\\
   &&=\sum_xe(x)
   f(x)^{T*}\sum_\mu\gamma^\mu(x)\{U(x,\mu)g(x+\hat\mu)-g(x)\}
\nonumber\\
   &&=\sum_{x,\mu}\Bigl\{
   e(x-\hat\mu)f(x-\hat\mu)^{T*}\gamma^\mu(x-\hat\mu)U(x-\hat\mu,\mu)
   -e(x)f(x)^{T*}\gamma^\mu(x)\Bigr\}g(x)
\nonumber\\
   &&=-\sum_{x,\mu}e(x-\hat\mu)
   \Bigl[U(x-\hat\mu,\mu)^{-1}\gamma^\mu(x-\hat\mu)
   U(x-\hat\mu,\mu)
\nonumber\\
   &&\qquad\qquad\qquad\qquad\qquad\qquad\qquad\quad
   \times
   \left\{f(x)-U(x-\hat\mu,\mu)^{-1}f(x-\hat\mu)\right\}\Bigr]^{T*}g(x)
\nonumber\\
   &&\qquad-\sum_{x,\mu}
   \left\{e(x)\gamma^\mu(x)f(x)
   -e(x-\hat\mu)U(x-\hat\mu,\mu)^{-1}\gamma^\mu(x-\hat\mu)
   U(x-\hat\mu,\mu)f(x)\right\}^{T*}g(x)
\nonumber\\
   &&\equiv-(\nablaslash^*f,g)
   -\sum_x
   \left\{\sum_\mu\nabla_\mu^*\{e(x)\gamma^\mu(x)\}f(x)\right\}^{T*}g(x),
\label{threexfour}
\end{eqnarray}
where, in the second equality, we have shifted the coordinate~$x$ to
$x-\hat\mu$. This manipulation is justified for a lattice with infinite
extent or for a finite size lattice with periodic boundary conditions. In
the last line, we have introduced a lattice Dirac operator defined in terms
of the backward covariant difference,
\begin{eqnarray}
   \nablaslash^*\psi(x)
   &\equiv&\sum_\mu
   e(x)^{-1}e(x-\hat\mu)
   U(x-\hat\mu,\mu)^{-1}\gamma^\mu(x-\hat\mu)U(x-\hat\mu,\mu)
\nonumber\\
   &&\qquad\qquad\qquad\qquad\qquad\quad\times
   \left\{\psi(x)-U(x-\hat\mu,\mu)^{-1}\psi(x-\hat\mu)\right\},
\label{threexfive}
\end{eqnarray}
and the covariant divergence of $e(x)\gamma^\mu(x)$,
\begin{eqnarray}
   &&\sum_\mu\nabla_\mu^*\{e(x)\gamma^\mu(x)\}
\nonumber\\
   &&
   \equiv
   \sum_\mu\{e(x)\gamma^\mu(x)
   -e(x-\hat\mu)U(x-\hat\mu,\mu)^{-1}\gamma^\mu(x-\hat\mu)
   U(x-\hat\mu,\mu)\}.
\label{threexsix}
\end{eqnarray}
Although its structure is somewhat complicated, $\nablaslash^*$ behaves
covariantly under the local Lorentz transformation~(\ref{twoxseven}) and
coincides with the continuum Dirac operator~$\Dslash$ in the naive continuum
limit.

We note that under the parametrization~(\ref{threextwo}), the continuum
limit of Eq.~(\ref{threexsix}) becomes
\begin{equation}
   \lim_{a\to0}\sum_\mu\nabla_\mu^*\{e(x)\gamma^\mu(x)\}=
   \partial_\mu\{e(x)\gamma^\mu(x)\}
   +e(x){1\over2}\omega_{\mu ab}(x)[\sigma^{ab},\gamma^\mu(x)],
\label{threexseven}
\end{equation}
which is the covariant divergence of $e(x)\gamma^\mu(x)$ in the
continuum, $\nabla_\mu\{e(x)\gamma^\mu(x)\}$. This combination vanishes
identically if the spin connection and the vielbein are related through the
metric condition. Thus, as a lattice counterpart of this property, we
postulate that the external vielbein and the link variables satisfy the
following constraint:
\begin{equation}
   \sum_\mu\nabla_\mu^*\{e(x)\gamma^\mu(x)\}=0,\qquad\hbox{for all $x$}.
\label{threexeight}
\end{equation}
The equality~(\ref{threexfour}) then shows that the
combination~$\nablaslash^*$ in eq.~(\ref{threexfive}) is precisely the opposite
of the hermitian conjugate of $\nablaslash$ in Eq.~(\ref{threexone}) with
respect to the inner product~(\ref{threexthree}). Using the
symbol~$\dagger$ to express this conjugation, we have\footnote{Our
result can also be obtained by using the ``weighted variables''
$\widetilde\psi(x)\equiv e(x)^{1/2}\psi(x)$,
$\widetilde{\overline\psi}(x)\equiv e(x)^{1/2}\overline\psi(x)$ and the
``naive'' inner product $(f,g)_{\rm naive}\equiv\sum_xf(x)^{T*}g(x)$. In that
case, the hermitian conjugation with respect to the naive inner product is just
$T*$. Then, re-expressing the Dirac operators acting on the weighted variables
as ones acting on the original variables, we obtain the expressions appearing
above.}
\begin{equation}
   \nablaslash^\dagger=-\nablaslash^*,\qquad
   (\nablaslash^*)^\dagger=-\nablaslash.
\label{threexnine}
\end{equation}

The structure of $\nablaslash^*$, defined in Eq.~(\ref{threexfive}), is more
complicated than that of $\nablaslash$, defined in Eq.~(\ref{threexone}), and,
partially due to this fact, any explicit analytical calculations involving our
Dirac operator are expected to become quite involved. Although our choice of
$\nablaslash$ and $\nablaslash^*$ given above is by no means unique, it turns
out that one or the other of $\nablaslash$ and $\nablaslash^*$ constituting
such a conjugate pair must be somewhat complicated, as for the choice we
consider. Noting this point, we adopt the above choice in this paper.

Having clarified the meaning of the hermitian conjugation, it is
straightforward to construct a lattice Dirac operator of overlap type
with the desired properties. We first define the Wilson-Dirac operator by
\begin{equation}
   D_{\rm w}={1\over2}
   \left\{\nablaslash+\nablaslash^*
   -{1\over2}(\nablaslash^*\nablaslash+\nablaslash\nablaslash^*)\right\},
\label{threexten}
\end{equation}
where the second term is the Wilson term in curved space, which is hermitian
with respect to the inner product~(\ref{threexthree}). This Wilson term
ensures the absence of species doubling in our overlap-type
Dirac operator constructed below. The naive continuum limit
of~$D_{\rm w}$ coincides with the continuum Dirac operator~$\Dslash$.
Due to the conjugation property~(\ref{threexnine}), we have the $\gamma_5$
hermiticity of the Wilson-Dirac operator
\begin{equation}
   D_{\rm w}^\dagger=\gamma_5D_{\rm w}\gamma_5,
\label{threexeleven}
\end{equation}
which is crucial in the following analysis.

\subsection{Overlap-Dirac operator, Ginsparg-Wilson relation and the lattice
index theorem}
In analogy to the case of lattice gauge theory, we define the overlap-Dirac
operator from the Wilson-Dirac operator~(\ref{threexten}) as
\begin{equation}
   D=1-A(A^\dagger A)^{-1/2},\qquad A\equiv1-D_{\rm w}.
\label{threextwelve}
\end{equation}
The naive continuum limit of the operator~$D$ is $\Dslash$. In the flat space
limit, in which we have $e_\mu^a(x)=\delta_\mu^a$ and $U(x,\mu)=1$, the free
Dirac operator in the momentum space reads
\begin{equation}
   \widetilde D(k)
   ={1\over a}
   -{{1\over a}-\sum_\mu(i\gamma^\mu\ring k_\mu+{1\over 2}a\hat k_\mu^2)
   \over[1+{1\over2}a^4\sum_{\mu<\nu}\hat k_\mu^2\hat k_\nu^2]^{1/2}},
\label{threexthirteen}
\end{equation}
where the lattice spacing~$a$ has been restored and we have
\begin{equation}
   \ring k_\mu\equiv{1\over a}\sin(ak_\mu),\qquad
   \hat k_\mu\equiv{2\over a}\sin\left({ak_\mu\over2}\right).
\label{threexfourteen}
\end{equation}
We thus see that $\widetilde D(k)^\dagger\widetilde D(k)=0$ implies~$k=0$,
and the free Dirac operator does not suffer from species doubling.

From its construction~(\ref{threextwelve}) and the $\gamma_5$ hermiticity
of the Wilson-Dirac operator~(\ref{threexeleven}), we find that the operator
satisfying the Ginsparg-Wilson relation in the conventional form,
\begin{equation}
   \gamma_5D+D\gamma_5=D\gamma_5D.
\label{threexfifteen}
\end{equation}
Or, in terms of the lattice-modified chiral
matrix~\cite{Narayanan:1998uu,Niedermayer:1998bi}
\begin{equation}
   \hat\gamma_5\equiv\gamma_5(1-D),\qquad
   (\hat\gamma_5)^\dagger=\hat\gamma_5,\qquad
   (\hat\gamma_5)^2=1,
\label{threexsixteen}
\end{equation}
the relation~(\ref{threexfifteen}) is expressed as
\begin{equation}
   D\hat\gamma_5=-\gamma_5D.
\label{threexseventeen}
\end{equation}
The Dirac operator~$D$ is also $\gamma_5$ hermitian, i.e.,
\begin{equation}
   D^\dagger=\gamma_5D\gamma_5,
\label{threexeighteen}
\end{equation}
with respect to the inner product~(\ref{threexthree}).\footnote{From this
$\gamma_5$ hermiticity, it follows that the Dirac determinant
$\int\prod_x\rmd\psi(x)\rmd\overline\psi(x)\,e^{-S_{\rm F}}$ is real.}

Thus we have obtained a lattice Dirac operator of overlap type that
describes the propagation of a single Dirac fermion in external
gravitational fields. Since the forms of the Ginsparg-Wilson relation and the
$\gamma_5$ hermiticity are identical to those in lattice gauge theory, we
can repeat the same argument for the index theorem in the latter theory with
slight modifications for curved space.

A natural lattice action for the massless Dirac fermion in curved space is
\begin{equation}
   S_{\rm F}=\sum_xe(x)\overline\psi(x)D\psi(x).
\label{threexnineteen}
\end{equation}
The invariance of this action under the local Lorentz transformation is
obvious. This action is also invariant under the modified chiral transformation
\begin{equation}
   \psi(x)\to\left(1+i\theta\hat\gamma_5\right)\psi(x),\qquad
   \overline\psi(x)\to\overline\psi(x)\left(1+i\theta\gamma_5\right)
\label{threextwenty}
\end{equation}
where $\theta$ is an infinitesimal constant parameter, due to the
Ginsparg-Wilson relation~(\ref{threexseventeen}). The functional
integration measure is, however, not invariant under this transformation and
gives rise to a non-trivial Jacobian~$J$:
\begin{equation}
   \prod_x\rmd\psi(x)\rmd\overline\psi(x)
   \to J\prod_x\rmd\psi(x)\rmd\overline\psi(x),
\label{threextwentyone}
\end{equation}
where
\begin{eqnarray}
   \ln J&=&
   -i\theta\sum_x\tr\left\{\hat\gamma_5(x,x)+\gamma_5\delta_{xx}\right\}
\nonumber\\
   &\equiv&-2i\theta\sum_x\tr\Gamma_5(x,x)
\label{threextwentytwo}
\end{eqnarray}
and
\begin{equation}
   \Gamma_5(x,y)\equiv\gamma_5\left(\delta_{xy}-{1\over2}D(x,y)\right).
\label{threextwentythree}
\end{equation}
The operator~$\Gamma_5$ anti-commutes with the hermitian operator~$H$,
again due to the Ginsparg-Wilson relation:
\begin{equation}
   \{H,\Gamma_5\}=0,\qquad H\equiv\gamma_5D,\qquad H^\dagger=H.
\label{threextwentyfour}
\end{equation}

We introduce the eigenfunctions of $H$ as
\begin{equation}
   H\varphi_r(x)=\lambda_r\varphi_r(x),\qquad\lambda_r\in\mathbb{R}
\label{threextwentyfive}
\end{equation}
and assume that the eigenfunctions are normalized with respect to the inner
product~(\ref{threexthree}):
\begin{equation}
   (\varphi_r,\varphi_r)=1,\qquad\hbox{for all~$r$}.
\label{threextwentysix}
\end{equation}
Then, by the hermiticity of~$H$ with respect to the inner product, we have
\begin{equation}
   (\varphi_r,\varphi_s)=0,\qquad\hbox{for $\lambda_r\neq\lambda_s$}.
\label{threextwentyseven}
\end{equation}
The completeness of eigenmodes is expressed as
\begin{equation}
   \sum_r\varphi_r(x)\varphi_r(y)^{T*}=\delta_{xy}e(y)^{-1}.
\label{threextwentyeight}
\end{equation}

Now we can prove the lattice index theorem. The quantity
\begin{eqnarray}
   \sum_x\tr\Gamma_5(x,x)&=&\sum_x\tr\sum_y\Gamma_5(x,y)\delta_{yx}
\nonumber\\
   &=&\sum_xe(x)\sum_r\varphi_r(x)^{T*}\sum_y\Gamma_5(x,y)\varphi_r(y)
\nonumber\\
   &=&\sum_r(\varphi_r,\Gamma_5\varphi_r)
\label{threextwentynine}
\end{eqnarray}
in the Jacobian~(\ref{threextwentytwo}) is an {\it integer\/} even for finite
lattice spacings (i.e., before taking the continuum limit). This can easily be
seen by noting that the eigenvalue of $\Gamma_5\varphi_r$ is $-\lambda_r$ and
thus $\Gamma_5\varphi_r$ is orthogonal to $\varphi_r$ when $\lambda_r\neq0$.
Hence only zero eigenmodes contribute to the sum over $r$ in
Eq.~(\ref{threextwentynine}). Noting finally that $\Gamma_5\to\gamma_5$ for
zero modes of~$H$ and the normalization~(\ref{threextwentysix}), we have
\begin{equation}
   \sum_x\tr\Gamma_5(x,x)=n_+-n_-,
\label{threexthirty}
\end{equation}
where $n_\pm$ denote the numbers of zero eigenmodes with positive and negative
chiralities, respectively. This is the lattice index theorem in the presence of
a gravitational field.

\subsection{Classical continuum limit of the lattice index density}
We next consider the classical continuum limit of the lattice
index~(\ref{threextwentynine}). We consider the classical continuum limit to be
the $a\to0$ limit in which derivatives of the external gravitational
fields are assumed to be $O(a^0)$.\footnote{Note that we cannot take the
{\it naive\/} continuum limit of $\Gamma_5$ in
Eq.~(\ref{threextwentynine}) {\it a priori}, because the function on which
$\Gamma_5$ acts (namely $\delta_{yx}$) varies rapidly in lengthscale~$a$.} It is
desirable to evaluate the continuum limit directly in the same way as in
lattice gauge
theory~\cite{Kikukawa:1998pd,Adams:1998eg,Suzuki:1998yz,Fujiwara:2002xh}, but
the actual calculation is quite involved. Instead, here we resort to a
powerful argument due to Fujikawa~\cite{Fujikawa:1998if}, which utilizes the
topological nature of the lattice index~(\ref{threextwentynine}). We
slightly modify the original argument to make the reasoning more transparent.

First we note that in the expressions of the
index~(\ref{threextwentynine}), $\Gamma_5$ can be replaced by
$\Gamma_5e^{-{H^2/M^2}}$ with an arbitrary mass~$M$, because only zero
eigenmodes of $H$ contribute to the index. Thus we have
\begin{equation}
   \sum_x\tr\Gamma_5(x,x)=\sum_x\tr\sum_y\Gamma_5e^{-H^2/M^2}(x,y)
   \delta_{yx}
   \equiv a^{2n}\sum_x{\mathcal A}_5(x),
\label{threexthirtyone}
\end{equation}
where the lattice spacing~$a$ has been restored, and the index density has
been defined by
\begin{equation}
   {\mathcal A}_5(x)\equiv
   {1\over a^{2n}}\tr\sum_y\Gamma_5e^{-H^2/M^2}(x,y)\delta_{yx}.
\label{threexthirtytwo}
\end{equation}
From this point, we consider the index density on a lattice with infinite
extent, as usual for the classical continuum limit. Then we have
\begin{eqnarray}
   {\mathcal A}_5(x)&=&\int_{\mathcal B}{\rmd^{2n}k\over(2\pi)^{2n}}\,
   e^{-ikx}\tr\sum_y\Gamma_5e^{-H^2/M^2}(x,y)e^{iky},
\nonumber\\
   &=&\int_{\mathcal B}{\rmd^{2n}k\over(2\pi)^{2n}}\,
   e^{-ikx}\tr\left(\gamma_5-{a\over2}H\right)e^{-{H^2/M^2}}e^{ikx},
\label{threexthirtythree}
\end{eqnarray}
where the Brillouin zone is denoted by ${\mathcal B}$:
\begin{equation}
   {\mathcal B}\equiv\left\{k\in{\mathbb R}^{2n}\Big|
   -{\pi\over a}<k_\mu\leq{\pi\over a}\right\}.
\label{threexthirtyfour}
\end{equation}

Now suppose that we expand the integrand of Eq.~(\ref{threexthirtythree})
with respect to the external fields $h_\mu^a(x)\equiv e_\mu^a(x)-\delta_\mu^a$,
$\omega_{\mu ab}(x)$ and their derivatives. Formally, this yields
\begin{eqnarray}
   &&e^{-ikx}\tr\left(\gamma_5-{a\over2}H\right)e^{-{H^2/M^2}}e^{ikx}
\nonumber\\
   &&=\sum_Ic_I(k;a,M)\,
   \partial^{\alpha_1}h_{\mu_1}^{a_1}(x)\cdots
   \partial^{\alpha_p}h_{\mu_p}^{a_p}(x)
   \partial^{\beta_1}\omega_{\nu_1b_1c_1}(x)\cdots
   \partial^{\beta_q}\omega_{\nu_qb_qc_q}(x),
\nonumber\\
\label{threexthirtyfive}
\end{eqnarray}
where the superscripts of the partial derivatives $\alpha_i$ and~$\beta_j$ are
used to label multiple partial derivatives [for example,
$\partial^\alpha=(\partial_0)^{\alpha_0}\cdots
(\partial_{2n-1})^{\alpha_{2n-1}}$], and $I$ denotes the collection of indices:
$I=\{\alpha_1,\ldots,\alpha_p;a_1,\ldots,a_p;
\beta_1,\ldots,\beta_q;
b_1,\ldots,b_q;c_1,\ldots,c_q\}$. One can then confirm, through some
examination, that the coefficient~$c_I(k;a,M)$ has the structure
\begin{eqnarray}
   &&c_I(k;a,M)=p_I(\hat k;k,a,M)
   \exp\left\{-{2\over a^2M^2}\left(1
   -{1-{1\over2}a^2\sum_\mu\hat k_\mu^2\over
   [1+{1\over2}a^4\sum_{\mu<\nu}\hat k_\mu^2\hat k_\nu^2]^{1/2}}
   \right)\right\},
\nonumber\\
\label{threexthirtysix}
\end{eqnarray}
where $p_I(\hat k;k,a,M)$ is {\it a polynomial in $\hat k_\mu$ whose
coefficients are bounded functions of $k\in{\mathcal B}$ and of $O(a^0)$}.
Noting this structure and the inequality
\begin{equation}
   {2\over\pi}|k_\mu|\leq|\hat k_\mu|\leq|k_\mu|,\qquad
   \hbox{for all $k\in{\mathcal B}$},
\label{threexthirtyseven}
\end{equation}
we infer that there exists a polynomial of $|k_\mu|$ with positive
coefficients, $b_I(k;M)$, and some positive number~$\varepsilon$
such that, for any $a\leq\varepsilon$, we have
\begin{equation}
   |c_I(k;a,M)|\leq b_I(k;M)
   \exp\left\{-{4\over\pi^2M^2}\sum_\mu k_\mu^2\right\},
   \qquad\hbox{for all $k\in{\mathcal B}$}.
\label{threexthirtyeight}
\end{equation}

Next, we write
\begin{equation}
   \int_{\mathcal B}\rmd^{2n}k\,c_I(k;a,M)
   =\int_\Delta\rmd^{2n}k\,c_I(k;a,M)
   +\int_{{\mathcal B}-\Delta}\rmd^{2n}k\,c_I(k;a,M),
\label{threexthirtynine}
\end{equation}
where $\Delta$ is a box of size~$\Lambda$ in the Brillouin zone
(from this point, the lattice spacing~$a$ is assumed to be smaller
than~$\varepsilon$)
\begin{equation}
   \Delta\equiv
   \left\{k\in{\mathbb R}^{2n}\mid|k_\mu|\leq\Lambda<\pi/\varepsilon\right\}.
\end{equation}
From the bound~(\ref{threexthirtyeight}), we have
\begin{equation}
   \left|\int_{{\mathcal B}-\Delta}\rmd^{2n}k\,c_I(k;a,M)\right|
   \leq\int_{{\mathcal B}-\Delta}\rmd^{2n}k\,
   b_I(k;M)\exp\left\{-{4\over\pi^2M^2}\sum_\mu k_\mu^2\right\},
\end{equation}
and thus
\begin{equation}
   \lim_{\Lambda\to\infty}\lim_{a\to0}
   \int_{{\mathcal B}-\Delta}\rmd^{2n}k\,c_I(k;a,M)=0,
\end{equation}
because $b_I(k;M)\exp\{-4\sum_\mu k_\mu^2/(\pi^2M^2)\}$ is integrable
in ${\mathbb R}^{2n}$:
\begin{equation}
   \int_{{\mathbb R}^{2n}}\rmd^{2n}k\,
   b_I(k;M)\exp\left\{-{4\over\pi^2M^2}\sum_\mu k_\mu^2\right\}<\infty.
\end{equation}
We thus conclude from Eq.~(\ref{threexthirtynine}) that
\begin{eqnarray}
   \lim_{a\to0}\int_{\mathcal B}\rmd^{2n}k\,c_I(k;a,M)
   &=&\lim_{\Lambda\to\infty}\lim_{a\to0}\int_\Delta\rmd^{2n}k\,c_I(k;a,M)
\nonumber\\
   &=&\int_{{\mathbb R}^{2n}}\rmd^{2n}k\,\lim_{a\to0}c_I(k;a,M).
\label{threexfortyfour}
\end{eqnarray}
That is, the $a\to0$ limit of the integral of $c_I(k;a,M)$ over ${\mathcal B}$
is given by an ${\mathbb R}^{2n}$ integration of the $a\to0$ limit of
$c_I(k;a,M)$. The latter should be regarded as the {\it naive\/} continuum
limit, because the momentum~$k$ carried by the plane wave~$e^{ikx}$ is kept
fixed in this $a\to0$ limit.

Now, by applying Eq.~(\ref{threexfortyfour}) to all terms of the
expansion~(\ref{threexthirtyfive}), we obtain
\begin{eqnarray}
   \lim_{a\to0}\mathcal{A}_5(x)
   &=&\int_{{\mathbb R}^{2n}}
   {\rmd^{2n}k\over(2\pi)^{2n}}\,\lim_{a\to0}
   e^{-ikx}\tr\left(\gamma_5-{a\over2}H\right)e^{-{H^2/M^2}}e^{ikx}
\nonumber\\
  &=&\int_{{\mathbb R}^{2n}}
   {\rmd^{2n}k\over(2\pi)^{2n}}\,
   e^{-ikx}\tr\gamma_5e^{\Dslash^2/M^2}e^{ikx},
\label{threexfortyfive}
\end{eqnarray}
because $\lim_{a\to0}H=\gamma_5\Dslash$ in the naive continuum limit. It is
interesting that the term proportional to $aH/2$ does not contribute in this
calculational scheme~\cite{Fujikawa:1998if}. In this way, we obtain the
index density {\it in the continuum theory}. The underlying important point
in our argument is that the lattice free Dirac operator does not possess
doubler's zero, and $e^{-H^2/M^2}$ acts as a suppression factor at the
boundary of the Brillouin zone for $a\to0$.

The calculation of the index density in the continuum (which can be
evaluated in the $M\to\infty$ limit) is
well-known~\cite{Alvarez-Gaume:1983at}. (For a calculation in the plane wave
basis, see Ref.~\citen{Fujikawa:1986hk}.) The result is given by the
so-called Dirac genus~\cite{Atiyah:1967,Hawking:1979zs,Alvarez-Gaume:1983at},
\begin{equation}
   \lim_{a\to0}a^{2n}\sum_x\tr\Gamma_5(x,x)
   =\int_{M_{2n}}\lim_{a\to0}{\mathcal A}_5(x)
   =\int_{M_{2n}}
   \det\left\{{i\hat R/4\pi\over\sinh(i\hat R/4\pi)}\right\}^{1/2},
\label{threexfortysix}
\end{equation}
where the curvature 2-form is defined by
\begin{equation}
   (\hat R)_a{}^b={1\over2}R_{\mu\nu a}{}^b\,\rmd x^\mu\wedge\rmd x^\nu,
\end{equation}
and the determinant, $\det$, is taken with respect to the Lorentz indices
of~$\hat R$. Thus as expected from the absence of the species doubling
(which implies the correct number of degrees of freedom) and the topological
properties of the lattice index, we find that our formulation reproduces the
correct expression of the axial $\U(1)$ anomaly in a gravitational field. This
demonstration can be regarded as a test of the restoration of the general
coordinate invariance in the classical continuum limit.

As noted in the introduction, the gauge invariance and the topological
nature of the axial anomaly in lattice gauge theory allow a
cohomological analysis of the anomaly~\cite{Luscher:1999kn}, as in the
continuum theory~\cite{Barnich:2000zw}. It might be possible to carry out this
kind of analysis in the present lattice formulation for the gravitational
interaction, because the index density~(\ref{threexthirtytwo}) is local,
Lorentz invariant and possesses a topological property.\footnote{For a
lattice with finite extent, $a^{2n}\sum_x\delta{\mathcal A}_5(x)=0$
clearly holds, from the index theorem~(\ref{threexthirty}), where $\delta$
denotes an arbitrary variation of gravitational fields. Even for a lattice with
infinite extent, this equation is meaningful, provided that the
variation~$\delta$ has finite support and the Dirac operator is local.
(We did not prove the latter for gravitational fields of finite strength.)
Then, this topological property can be shown directly from the
Ginsparg-Wilson relation for any $M$.} The absence of a manifest general
coordinate invariance in our lattice formulation, however, could be an
obstacle for such an analysis.

\subsection{Reduction to the Majorana fermion}
We now briefly comment on a reduction of the Dirac fermion to the Majorana
fermion with our lattice Dirac operator. We can apply the prescription in
Euclidean field theory, which is reviewed in Ref.~\citen{Inagaki:2004ar} also
to the present case of a gravitational interaction. Naively, one expects that
the Majorana fermion can be defined in $8k$, $8k+2$ and $8k+4$ dimensions (and
in the odd number of dimensions $8k+1$ and $8k+3$), because the spinor
representation of the Lorentz group is real for these number of dimensions.

The prescription starts by setting
\begin{equation}
   \psi={1\over\sqrt{2}}(\chi+i\eta),\qquad
   \overline\psi={1\over\sqrt{2}}(\chi^TB-i\eta^TB)
\label{threexfortyeight}
\end{equation}
in Eq.~(\ref{threexnineteen}), where $B$ denotes the charge conjugation matrix,
either $B_1$ or $B_2$. (We shall use the notation of
Ref.~\citen{Inagaki:2004ar}). From the definition~(\ref{threextwelve}) and
properties of~$B$~\cite{Inagaki:2004ar}, it is straightforward to confirm the
skew-symmetricity expressed by
\begin{equation}
   e(y)(BD)^T(y,x)=-e(x)BD(x,y),
\label{threexfortynine}
\end{equation}
where the transpose $T$ acts on the spinor indices, with $B=B_1$ for $8k+2$
dimensions and $B=B_2$ for $8k+4$ dimensions. (To show this, we have to
use the constraint~(\ref{threexeight}).) Due to this skew-symmetric property,
the action for the Dirac fermion~(\ref{threexnineteen}) decomposes into two
pieces, describing mutually independent systems,
\begin{equation}
   S_{\rm F}={1\over2}\sum_xe(x)\chi^T(x)BD\chi(x)
   +{1\over2}\sum_xe(x)\eta^T(x)BD\eta(x),
\end{equation}
for $8k+2$ and $8k+4$ dimensions. Thus, taking either of these two pieces as the
action, say,
\begin{equation}
   S_{\rm M}={1\over2}\sum_xe(x)\chi^T(x)BD\chi(x),
\end{equation}
we can define the Majorana fermion in a gravitational field as half of the
Dirac fermion. Moreover, by repeating the argument of
Ref.~\citen{Inagaki:2004ar}, we can show that the partition function of the
Majorana fermion (which is manifestly local and Lorentz invariant),
\begin{equation}
   \Pf\{eBD\}=\int\prod_x\rmd\chi(x)\,e^{-S_{\rm M}},
\end{equation}
is semi-positive definite.

However, this prescription does not work for $8k$~dimensions, because our
lattice Dirac operator does not possess the property~(\ref{threexfortynine})
for use of either $B_1$ or $B_2$. (This is basically due to the presence of the
Wilson term.) This is expected, as argued in Ref.~\citen{Inagaki:2004ar},
because a single Majorana fermion in $8k$~dimensions may suffer from
global gravitational/gauge anomalies~\cite{Alvarez-Gaume:1983ig}.
Specifically, if
there exists a simple local Lorentz invariant lattice formulation of the
Majorana fermion in $8k$~dimensions, it would not reproduce the global
gravitational/gauge anomalies.\footnote{For odd number of dimensions, one can
try the prescription~(\ref{threexfortyeight}) with the Wilson-Dirac
operator~(\ref{threexten}). It is then found that the skew-symmetric property
$(B_1D_{\rm w})^T(y,x)e(y)=-e(x)B_1D_{\rm w}(x,y)$ holds for $8k+3$
dimensions, but not for $8k+1$ dimensions. The global anomaly exists in $8k+3$
dimensions~\cite{Alvarez-Gaume:1983ig,Hayakawa:2006fd}, and this is
consistent with the above observation.}

\section{Other Lorentz representations}
We now consider a generalization of our construction to systems with
matter fields in other Lorentz (reducible) representations. This is easily
achieved. The link variable in the vector representation is given by
\begin{equation}
   {\mathcal U}_a{}^b(x,\mu)\equiv{1\over2^n}
   \tr\{\gamma_aU(x,\mu)\gamma^bU(x,\mu)^{-1}\}\in\SO(2n),
\label{fourxone}
\end{equation}
which reads ${\mathcal U}_a{}^b(x,\mu)=\delta_a^b+\omega_{\mu a}{}^b(x)
+\cdots$ with the parametrization~(\ref{threextwo}). For example, the
lattice Dirac operators for the spinor-vector field~$\psi_a(x)$ can be
defined by
\begin{eqnarray}
   &&\nablaslash\psi_a(x)\equiv\sum_\mu
   \gamma^\mu(x)\{U(x,\mu)\sum_b{\mathcal U}_a{}^b(x,\mu)\psi_b(x+\hat\mu)
   -\psi_a(x)\},
\nonumber\\
   &&\nablaslash^*\psi_a(x)\equiv\sum_\mu
   e(x)^{-1}e(x-\hat\mu)
   U(x-\hat\mu,\mu)^{-1}\gamma^\mu(x-\hat\mu)U(x-\hat\mu,\mu)
\nonumber\\
   &&\qquad\qquad\qquad\qquad
   \times\{\psi_a(x)
   -U(x-\hat\mu,\mu)^{-1}\sum_b{\mathcal U}^{-1}{}_a{}^b(x-\hat\mu,\mu)
   \psi_b(x-\hat\mu)\},
\label{fourxtwo}
\end{eqnarray}
and these operators are conjugate to each other with respect to the following
natural inner product for the spinor-vector fields:
\begin{equation}
   (f,g)_{\rm sv}\equiv\sum_x e(x)\sum_af_a(x)^{T*}g_a(x).
\end{equation}
Thus the overlap-Dirac operator for the spinor-vector fields can be defined
by combinations analogous to Eqs.~(\ref{threextwelve}) and~(\ref{threexten}).
The lattice action is given by
\begin{equation}
   S_{\rm sv}=\sum_xe(x)\sum_a\overline\psi_a(x)D\psi_a(x).
\label{fourxfour}
\end{equation}
Then, the argument for the axial anomaly and the lattice index proceeds in the
same manner as in \S3. The continuum limit is easily obtained because,
as is clear from Eq.~(\ref{fourxtwo}), the link variables in the vector
representation~${\mathcal U}_a{}^b$ can be regarded as gauge fields
associated with an internal $\SO(2n)$ gauge symmetry. Thus its contribution to
the index is given by the Chern character (for a direct calculation with the
overlap-Dirac operator, see Ref.~\citen{Fujiwara:2002xh}), and we have
\begin{equation}
   \lim_{a\to0}\sum_x\tr\Gamma_5(x,x)
   =\int_{M_{2n}}
   \det\left\{{i\hat R/4\pi\over\sinh(i\hat R/4\pi)}\right\}^{1/2}
   \tr\{e^{-i\hat R/(2\pi)}\}.
\label{fourxfive}
\end{equation}
The last trace $\tr$ is defined with respect to the Lorentz indices
of~$\hat R$. This is the well-known expression of the index for the
spinor-vector
representation~\cite{Atiyah:1967,Hawking:1979zs,Alvarez-Gaume:1983at}.
Here, we do not claim that we have constructed a lattice action of the
Rarita-Schwinger spin $3/2$ field. The action for the spinor-vector
field~(\ref{fourxfour}), however, might be regarded as a gauge-fixed form of
the action of the Rarita-Schwinger field, when supplemented with appropriate
ghost fields.

Similarly, for the bi-spinor field~$\psi(x)$, which has two spinor indices,
we may take
\begin{eqnarray}
   &&\nablaslash\psi(x)\equiv\sum_\mu
   \gamma^\mu(x)\{U(x,\mu)\psi(x+\hat\mu)U(x,\mu)^{-1}-\psi(x)\},
\nonumber\\
   &&\nablaslash^*\psi(x)\equiv\sum_\mu
   e(x)^{-1}e(x-\hat\mu)
   U(x-\hat\mu,\mu)^{-1}\gamma^\mu(x-\hat\mu)U(x-\hat\mu,\mu)
\nonumber\\
   &&\qquad\qquad\qquad\qquad\qquad
   \times\{\psi(x)
   -U(x-\hat\mu,\mu)^{-1}\psi(x-\hat\mu)U(x-\hat\mu,\mu)\}.
\label{fourxsix}
\end{eqnarray}
It can be verified that these two operators are conjugate to each other with
respect to a natural inner product for the bi-spinor field,
\begin{equation}
   (f,g)_{\rm bs}\equiv\sum_xe(x)\tr\{f(x)^{T*}g(x)\},
\label{fourxseven}
\end{equation}
if the constraint for the external fields, Eq.~(\ref{threexeight}) is
satisfied. The construction of the overlap-operator can be repeated, and the
action for the bi-spinor field is given by
\begin{equation}
   S_{\rm bs}=\sum_xe(x)\tr\{\overline\psi(x)D\psi(x)\}.
\label{fourxeight}
\end{equation}
As is clear from Eq.~(\ref{fourxsix}), the link variables on the right-hand
side of $\psi$ (which act on one of the spinor indices of the bi-spinor) can be
regarded as gauge fields associated with an internal $\mathop{\rm spin}(2n)$
gauge symmetry. Thus, the continuum limit of the index\footnote{Here, the
$\gamma_5$ chirality is understood to be defined with respect to the left
spinor index of the bi-spinor.} is obtained from Eq.~(\ref{fourxfive}) by
considering the Chern character in the spinor representation, and we have
\begin{equation}
   \lim_{a\to0}\sum_x\tr\Gamma_5(x,x)
   =\int_{M_{2n}}
   \det\left\{{2i\hat R/4\pi\over\tanh(i\hat R/4\pi)}\right\}^{1/2}.
\end{equation}
This is also the well-known expression of the index for the bi-spinor
representation~\cite{Atiyah:1967,Hawking:1979zs,Alvarez-Gaume:1983at}. On
the basis of the correspondence of the bi-spinor field with a collection of
anti-symmetric fields (differential forms), this formulation could be useful
for defining anti-symmetric tensor fields coupled to gravity on a lattice.

\section{Matter fields with a definite chirality}
Finally, we briefly consider the chirality projection with our overlap-type
lattice Dirac operator. For simplicity, we treat only the Weyl fermion.
Generalization to other cases, such as a spinor-vector or a bi-spinor
with a definite chirality, is straightforward. We follow the formulation given
in Ref.~\citen{Luscher:1999du} for lattice chiral gauge theories.

As in lattice gauge theory, we introduce chirality projection operators in the
forms
\begin{equation}
   \hat P_\pm={1\over2}(1\pm\hat\gamma_5),\qquad
   P_\pm={1\over2}(1\pm\gamma_5),
\end{equation}
where $\hat\gamma_5$ is given in Eq.~(\ref{threexsixteen}). The
chirality for the fermion is asymmetrically imposed as $\hat P_-\psi=\psi$
and $\overline\psi P_+=\overline\psi$. The partition function of the Weyl
fermion in an external gravitational field is then given by
\begin{equation}
   \int\rmD[\psi]\rmD[\overline\psi]\,e^{-S_{\rm F}}.
\end{equation}
(The lattice action $S_{\rm F}$ takes a form identical to that in
Eq.~(\ref{threexnineteen}).) We next introduce basis vectors $v_j(x)$, $j=1$, 2,
3, \dots, which satisfy
\begin{equation}
   \hat P_-v_j=v_j,\qquad\sum_xe(x)v_k(x)^{T*}v_j(x)=\delta_{kj},
\end{equation}
and $\overline v_k(x)$, $k=1$, 2, 3, \dots, which satisfy
\begin{equation}
   \overline v_kP_+=\overline v_k,
   \qquad\sum_xe(x)\overline v_k(x)\overline v_j(x)^{T*}=\delta_{kj}.
\end{equation}
Then we expand the Weyl fermion in terms of these bases as
$\psi(x)=\sum_jv_j(x)c_j$ and
$\overline\psi(x)=\sum_k\overline c_k\overline v_k(x)$. The functional
integration measure is then defined by $\rmD[\psi]\rmD[\overline\psi]\equiv
\prod_j\rmd c_j\prod_k\rmd\overline c_k$. The partition function is thus
given by $\det M$, where the matrix $M$ is
\begin{equation}
   M_{kj}\equiv\sum_xe(x)\overline v_k(x)Dv_j(x).
\end{equation}

Because covariance under the local Lorentz transformation is manifest in
our construction, it is easy to study how the partition function changes
under the local Lorentz transformation; this provides a lattice
counterpart of the local Lorentz anomaly~\cite{Chang:1984ib}, which is one
facet~\cite{Bardeen:1984pm} of the gravitational
anomalies~\cite{Alvarez-Gaume:1983ig,Alvarez-Gaume:1984dr}. Under
an infinitesimal local Lorentz transformation,
\begin{equation}
   \delta_\eta U(x,\mu)=\Theta(x)U(x,\mu)-U(x,\mu)\Theta(x+\hat\mu),
\end{equation}
where $\Theta(x)={1\over2}\sum_{ab}\Theta_{ab}(x)\sigma^{ab}$, the Dirac
operator behaves covariantly, as $\delta_\eta D=[\Theta,D]$. Then, noting the
completeness relations\footnote{In the first expression, the argument of
$e^{-1}$ must be $y$. This is consistent with the facts that the
``conjugate'' of a basis vector~$v_j$ satisfies the constraint
$\sum_yv_j(y)^{T*}e(y)\hat P_-(y,z)e(z)^{-1}=v_j(z)^{T*}$ and that
$(\hat P_-)^2=\hat P_-$.}
\begin{eqnarray}
   \sum_jv_j(x)v_j(y)^{T*}=\hat P_-(x,y)e(y)^{-1},\qquad
   \sum_k\overline v_k(x)^{T*}\overline v_k(y)=P_+\delta_{xy}e(y)^{-1},
\end{eqnarray}
we have
\begin{eqnarray}
   \delta_\eta\ln\det M
   &=&\sum_x\tr\{\Theta(x)\Gamma_5(x,x)\}
   +\sum_j\sum_xe(x)v_j(x)^{T*}\delta_\eta v_j(x)
\nonumber\\
   &\equiv& i\sum_x{1\over2}\sum_{ab}\Theta_{ab}(x){\mathcal A}^{ab}(x)
   -i\mathfrak{L}_\eta,
\label{fivexeight}
\end{eqnarray}
where
\begin{eqnarray}
   {\mathcal A}^{ab}(x)\equiv-i\tr\{\sigma^{ab}\Gamma_5(x,x)\},\qquad
   \mathfrak{L}_\eta\equiv i\sum_j(v_j,\delta_\eta v_j).
\end{eqnarray}
The first part, ${\mathcal A}^{ab}(x)$, which is covariant under the local
Lorentz transformation, corresponds to the covariant
form~\cite{Fujikawa:1983bg,Bardeen:1984pm} of the local Lorentz anomaly. The
second part, $\mathfrak{L}_\eta$, is the so-called measure
term~\cite{Luscher:1999du}, which parametrizes the manner in which the basis
vectors change
under a variation of a gravitational field (which in the present context is a
variation of an infinitesimal local Lorentz transformation). The measure term
cannot be completely arbitrary, because it is subject to the integrability
condition~\cite{Luscher:1999du}, and in Eq.~(\ref{fivexeight}), it provides
the lattice counterpart of (the divergence of) the Bardeen-Zumino
current~\cite{Bardeen:1984pm}, which gives rise to a difference between the
covariant form and the consistent form~\cite{Wess:yu,Bardeen:1984pm} of
anomalies. For the local Lorentz invariance to hold, basis
vectors for which $\delta_\eta\det M=0$ must be used. Whether this is possible
is the central question~\cite{Luscher:1999du,Luscher:1999kn}. In any case, the
cancellation of the local Lorentz anomaly in the continuum is necessary for
this anomaly cancellation in the lattice theory.

The covariant anomaly~${\mathcal A}^{ab}$ in our lattice formulation is given
by the expression of the axial $\U(1)$ anomaly~(\ref{threextwentytwo}) with
an additional generator of the gauge transformation~$\sigma^{ab}$ inserted.
This correspondence of the two anomalies also exists in the continuum theory.
The evaluation of the classical continuum limit of ${\mathcal A}^{ab}(x)$,
however, can be quite involved, because a lattice sum of ${\mathcal A}^{ab}(x)$
is not a topologically invariant quantity, and
we cannot apply the argument of Ref.~\citen{Fujikawa:1998if}. We do not study
this problem in the present paper.

The local Lorentz anomaly is only one facet of the gravitational
anomaly~\cite{Bardeen:1984pm}, and it is closely related to the anomaly
in the general coordinate invariance~\cite{Alvarez-Gaume:1983ig}. However, a
treatment of this anomaly in our lattice formulation, unfortunately, would
not be so simple, because the covariance (or invariance) under general
coordinate transformations is not manifest in our formulation. This is also a
subject of future study.

A more ambitious program employing the present formulation is to consider the
global gravitational anomalies~\cite{Alvarez-Gaume:1983ig,Witten:1985xe},
as done in Ref.~\citen{Bar:2000qa} for the global gauge
anomaly~\cite{Witten:fp} in lattice gauge theory.

\section{Conclusion}
In this paper, we constructed a lattice Dirac operator of overlap type
that describes the propagation of a single Dirac fermion in an external
gravitational field. The operator satisfies the conventional Ginsparg-Wilson
relation~(\ref{threexfifteen}) and possesses the $\gamma_5$
hermiticity~(\ref{threexeighteen}) with respect to the inner
product~(\ref{threexthree}). The lattice index theorem holds, as in lattice
gauge theory, and the classical continuum limit of the index density
reproduces the Dirac genus. Thus, the situation is, as far as the $\gamma_5$
symmetry and associated anomalies are concerned, analogous to the
case of lattice gauge theory.

However, there remain many points to be clarified. The first is to determine a
sufficient condition for the Dirac operator with gravitational fields of finite
strength to be local; this is known as the admissibility in the case of
lattice gauge theory~\cite{Hernandez:1999et}. To test the restoration of the
general coordinate invariance is another important problem. We observed such
restoration in a computation of the classical continuum limit of the lattice
index. This would not be a good example, however, because the restoration can be
guessed from the topological properties and the local Lorentz invariance. One
interesting test is to examine the trace anomaly, because its correct form is
fixed only when the general coordinate invariance is imposed.


\section*{Acknowledgements}
H.~S. would like to thank Takanori Fujiwara and Ryusuke Endo for preliminary
discussions that preceded the present work. We decided to seriously consider the
gravitational interaction of the overlap lattice fermion during the YITP
workshop ``Actions and Symmetries in Lattice Gauge Theory'' (YITP-W-05-25).
We would like to thank David B. Kaplan for his presentation at that workshop,
which partially motivated the present work, and the member of the Yukawa
Institute for Theoretical Physics at Kyoto University for their hospitality.
This work is supported in part by MEXT Grants-in-Aid for
Scientific Research (Nos.~13135203, 13135223, 17540242 and~17043004).

\end{document}